# Agentic AI in 6G Software Businesses: A Layered Maturity Model

Muhammad Zohaib
*Department of Software Engineering*
*Lappeenranta-Lahti University of Technology*
Lappeenranta, Finland
muhammad.zohaib@lut.fi

Muhammad Azeem Akbar
*Department of Software Engineering*
*Lappeenranta-Lahti University of Technology*
Lappeenranta, Finland
azeem.akbar@lut.fi

Sami Hyrynsalmi
*Department of Software Engineering*
*Lappeenranta-Lahti University of Technology*
Lappeenranta, Finland
Sami.Hyrynsalmi@lut.fi

Arif Ali Khan
*M3S Empirical Software Engineering Research Unit*
*University of Oulu*
Oulu, Finland
arif.khan@oulu.fi

*Abstract*— The emergence of agentic AI systems in 6G software businesses presents both strategic opportunities and significant challenges. While such systems promise increased autonomy, scalability, and intelligent decision-making across distributed environments, their adoption raises concerns regarding technical immaturity, integration complexity, organizational readiness, and performance-cost trade-offs. In this study, we conducted a preliminary thematic mapping to identify factors influencing the adoption of agentic software within the context of 6G. Drawing on a multivocal literature review and targeted scanning, we identified 29 motivators and 27 demotivators, which were further categorized into five high-level themes in each group. This thematic mapping offers a structured overview of the enabling and inhibiting forces shaping organizational readiness for agentic transformation. Positioned as a feasibility assessment, the study represents an early phase of a broader research initiative aimed at developing and validating a layered maturity model grounded in CMMI model with the software architectural three dimensions possibly Data, Business Logic, and Presentation. Ultimately, this work seeks to provide a practical framework to help software-driven organizations assess, structure, and advance their agent-first capabilities in alignment with the demands of 6G.

## I. INTRODUCTION

The transition to Sixth Generation (6G) networks is not merely a technological upgrade but a radical reimagination of the digital landscape, where software plays a central role as an intelligent, adaptive, and autonomous actor in distributed systems. As 6G shifts from performance-driven goals—such as ultra-low latency and high bandwidth toward service-driven paradigms like ubiquitous intelligence, semantic communication, and immersive experiences, software must evolve to meet unprecedented demands in autonomy, agility, and decision-making capability[1, 2]. This evolution compels a departure from traditional automation frameworks toward what this study refers to as agentic software: systems composed of self-directed agents capable of perception, reasoning, and collaboration within dynamically changing environments.

In this emerging paradigm, agentic software will form the core enabler of 6G use cases such as autonomous industry, context-aware smart cities, adaptive eHealth, and AI-enhanced edge computing [3]. These agents are designed to interact with each other (A2A), make goal-oriented decisions in real time, and adapt to environmental feedback with minimal human intervention[4]. The shift from orchestration-centric DevOps to distributed intelligence calls for new principles of software design, implementation, and lifecycle management that can support autonomy, scalability, and dynamic orchestration at scale[5].

Despite its growing relevance, the adoption of agentic software within 6G software businesses remains underexplored and poorly supported. Existing software maturity frameworks such as Capability Maturity Model Integration (CMMI) focus on process standardization and incremental improvements but lack constructs for reasoning-driven autonomy, multi-agent collaboration, or runtime adaptation[4] [3]. Similarly, current AI maturity models often emphasize data readiness, governance, or adoption of tooling, with little regard for the architectural and operational challenges inherent in agentic systems [6]. As a result, software businesses face a strategic dilemma: how to responsibly transition to agentic platforms while ensuring operational stability, business alignment, and long-term scalability.

Moreover, the research community has yet to articulate a structured model that helps organizations evaluate their readiness for agentic software adoption or navigate the complex trade-offs involved in building agent-based architectures for 6G environments. While frameworks for Agent-Oriented Software Engineering (AOSE) exist, they are primarily focused on design-time modeling and lack organizational maturity dimensions, transition guidelines, or empirical grounding in business settings [6].

To address this urgent gap, this paper proposes the development of an Agentic Software Maturity Model, a visionary, multi-dimensional framework designed to guide software-driven businesses in their progression from static automation to intelligent, self-governing software systems. Grounded in maturity model theory, agentic system architecture, and empirical software engineering literature, this Model will define levels of maturity based on agentic capability dimensions such as autonomy, inter-agent collaboration, and goal-driven orchestration. In doing so, it aims to provide a structured pathway for capability building, organizational adaptation, and business transformation in line with 6G objectives.

The obtain the objectives of this research our goal is to develop and validate a maturity model based on the foundational three-layer architecture (Data, Business Logic, and Presentation) that enables 6G software businesses to evaluate and advance their agent first capabilities. The research questions are outlined below.

- **RQ1:** What are the core challenges, success factors, and best practices in transitioning to agentic-first software systems in 6G software businesses?
- **RQ2:** How are these factors distributed and differentiated across the Data, Business Logic, and Presentation layers of the software architecture?
- **RQ3:** How can the identified factors be synthesized into a structured, multi-level maturity model?
- **RQ4:** How effective is the proposed layered maturity model in assessing and guiding real-world organizational transformation in agentic 6G software businesses?

The contributions of this paper are fourfold: (1) it defines the conceptual space of agentic software in the 6G context; (2) it identifies critical challenges and enablers through thematic synthesis of existing literature; (3) it proposes a layered maturity model aligned with business transformation trajectories; and (4) it outlines a roadmap for empirical validation and adoption in real-world software organizations. This visionary work serves as a foundational step toward equipping the software engineering community with the tools and frameworks necessary to operationalize agentic software for the 6G era.

### A. Motivation and Context of the Study

The transition to 6G presents software businesses with an inflection point in both technical capability and organizational strategy. While the technical vision of 6G includes ultra-low latency, sub-millisecond communication, distributed edge computing, and embedded AI-driven decision-making achieving these goals demands a fundamental rethinking of how software is developed, deployed, and maintained [1, 2]. Traditional DevOps pipelines based on manual orchestration, static workflows, and centralized architecture are misaligned with the core requirements of 6G, particularly in scenarios involving extreme heterogeneity, mobility, and autonomy [3, 5].

In this emerging landscape, agentic software systems offer potential to overcome critical limitations [7]. They embed reasoning and decision-making within autonomous agents that plan, act, and interact dynamically with humans and other agents [7]. Enabled by large language models and orchestration tools like LangChain, AutoGen, and Dapr, these agents integrate APIs, adapt to real-time inputs, and operate across distributed services [8, 9].

From a business perspective, the benefits of adopting agentic software for 6G are compelling. These include:

- **Cost reduction** through intelligent automation of workflows and incident handling,
- **Enhanced scalability** via distributed agent orchestration across edge and cloud environments,
- **Customization and adaptability** driven by real-time context-aware decision-making, and
- **Sustainability** through reduced reliance on human-in-the-loop processes and minimized operational latency.

Despite these advantages, widespread adoption remains elusive due to multiple uncertainties. First, the lack of standardized design and governance models makes agentic systems difficult to integrate into existing software ecosystems. Second, businesses are often unclear about the readiness criteria and cost-benefit trade-offs associated with deploying such systems. Third, the lack of strategic guidance in the form of maturity frameworks or best practices leaves decision-makers without tools to assess organizational readiness or measure progress over time [10].There is a clear need for early, theory-informed exploration into the factors that influence adoption, resistance, and strategic alignment of agentic software systems.

This motivates our present work. By identifying these underlying motivators and demotivators, and framing them within a visionary business readiness roadmap, we aim to fill a critical gap lack of maturity frameworks to guide the adoption of agentic AI in software businesses operating under 6G requirements (ultra-low latency, edge AI, extreme heterogeneity) in the software engineering discourse surrounding 6G.

## II. RESEARCH METHODOLOGY

This study identifies a preliminary set of motivators and demotivators related to the adoption of agentic systems in 6G software businesses. The overall research methodology is presented in Fig. 1.

This paper marks the planning phase of a broader research agenda. The goal is to explore and refine the key factors required to develop a maturity model for agentic software development. In the next phases, we will build a conceptual framework and validate it through empirical studies, including expert interviews and industrial case studies.The foundation of this work is grounded in our earlier multivocal literature review (MLR), which examined 133 sources including 102 peer-reviewed formal and 31 grey literature focused on 6G software business challenges [16]. That study followed the guidelines of Kitchenham and Charters [11] and Garousi et al. [12].

However, the MLR was intentionally broad in scope. During the analysis, several recurring challenges emerged—particularly those related to distributed orchestration, intelligent automation, and cost scalability. These pointed toward a promising but underexplored direction: agentic software development.

To further investigate this space, we conducted a targeted exploration using Google Scholar between April and June 2025. The search focused on the intersection of agentic software concepts with 6G technologies.

We applied three inclusion criteria for selecting studies. First, the documents had to be published after 2022. Second, they needed to be either formally peer-reviewed or recognized as industry-relevant grey literature. Third, each study had to be relevant to software adoption, business maturity, or agentic system deployment.

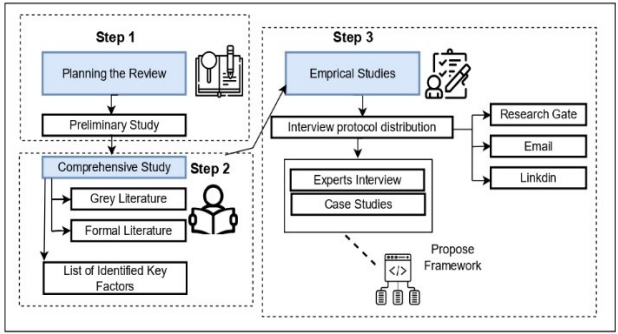

Fig. 1. Research Methodology.

Based on these criteria, a total of 46 documents were initially identified. After abstract screening and relevance assessment, 29 studies were selected for detailed analysis[11]. The findings were coded using both inductive and deductive approaches, following Braun and Clarke's six-phase thematic analysis protocol[13].

This study is exploration and conceptual by design. While the themes are grounded in validated MLR findings and refined using transparent search and analysis protocols, they should not be interpreted as exhaustive or empirically generalized. Future work will extend this vision by conducting expert interviews, survey-based validation, and case study evaluation to substantiate and refine the proposed maturity constructs.

### III. PRELIMINARY RESULTS

To explore how Agentic AI will being adopted in 6G software businesses. After selecting the relevant literature, we carefully reviewed each paper. Our goal was to find out what factors influence the use of Agentic AI in 6G software landscape. The identified motivators and depositors are presented as below.

#### A. Agentic AI Software in 6G: Motivator

As 6G ecosystems evolve toward intelligent, autonomous software architectures, agentic systems are emerging as a key enabler of this transformation. Our thematic analysis, based on 29 initial codes and 17 sub-themes, resulted in five core motivator themes, which are described below and visually presented in Fig. 2.

**M1:Scalable Autonomy** serves as a foundational driver of agentic system adoption. At its core, agent-based systems promote decentralized coordination and negotiation, where autonomous agents self-assign tasks without centralized control, improving robustness and fault tolerance [14][4][15]. Moreover, swarm-based dynamic processing enables agents to form adaptive clusters for distributed task execution[16-18],while fault-tolerant stateless execution supports rapid failover and self-recovery[19, 20]. Through edge-native autonomous decision-making, agents can operate close to data sources, reducing latency and improving responsiveness in time-critical contexts[21, 22]. Finally, autonomous runtime scalability allows agents to dynamically adjust resource utilization across microservices[19, 23], aligning with 6G's fluid, decentralized environments[15].

**M2:Cost Efficiency** is another central motivator, enabling leaner operational models. Agentic systems allow for low-code and no-code automation, where LLMs generate infrastructure logic with minimal human input [24]. With reusable design and templating[25], agent templates reduce redundancy and speed up deployment cycles[26-28]. Importantly, human workforce reduction is achieved through self-monitoring, conversational agents, and automated recovery, minimizing the need for 24/7 manual support [29, 30]. Lastly, cross-system deployment efficiency is enhanced as agents streamline integration across platforms, reducing multi-system coordination costs[24, 28, 31, 32].

**M3:Adaptive Intelligence** reflects the cognitive advantage offered by LLM-integrated agents. Real-time adaptive decision-making allows agents to interpret telemetry data and respond to changing conditions with context-aware prompts [24, 33, 34]. Meanwhile, LLM-driven behavior evolution enables agents to refine task logic dynamically based on performance feedback or environmental shifts [26, 27], supporting continuous optimization in high-variability systems.

**M4:Alignment with 6G Architecture** highlights how agentic design principles complement key 6G technical goals. With latency-aware edge deployment, agents are co-located at edge nodes to minimize communication delays and enable real-time services [26, 34]. 6G compliance alignment ensures agent frameworks meet URLLC and mMTC requirements, while remaining lightweight and bandwidth-efficient [3, 33, 34]. Additionally, containerized architectural conformance allows agent logic to be modular and microservice-ready [35] enabling plug-and-play composition with existing infrastructure layers.

Finally, **M5: Innovation & Differentiation** emphasizes how agentic systems create strategic value. Through customizable agent-based service logic, businesses can build proprietary workflows that enhance service personalization or automation [36, 37]. Autonomy as market differentiator enables companies to position their offerings as intelligent, adaptive, and AI-driven [38]. In summary, agentic systems align well with the operational demands and innovation strategies of 6G software businesses. Their ability to decentralize intelligence, reduce human and technical overhead, adapt in real-time, and integrate seamlessly with complex environments makes them a compelling path forward for organizations seeking agility, resilience, and future-readiness in the 6G era.

#### B. Agentic AI Software in 6G:Demotivators

Despite the promise of agentic systems for 6G software businesses, 27 pressing concerns hinder their adoption. Through thematic analysis, 17 sub-themes were synthesized into five high-level demotivator themes, which are described in detail below and illustrated in Fig. 3.

**D1: Technical Immaturity** remains one of the most foundational barriers. Many agentic frameworks suffer from a lack of runtime and API maturity[27, 30], marked by inconsistent interface standards and frequent changes in orchestration logic that impair reproducibility and interoperability[14, 38].

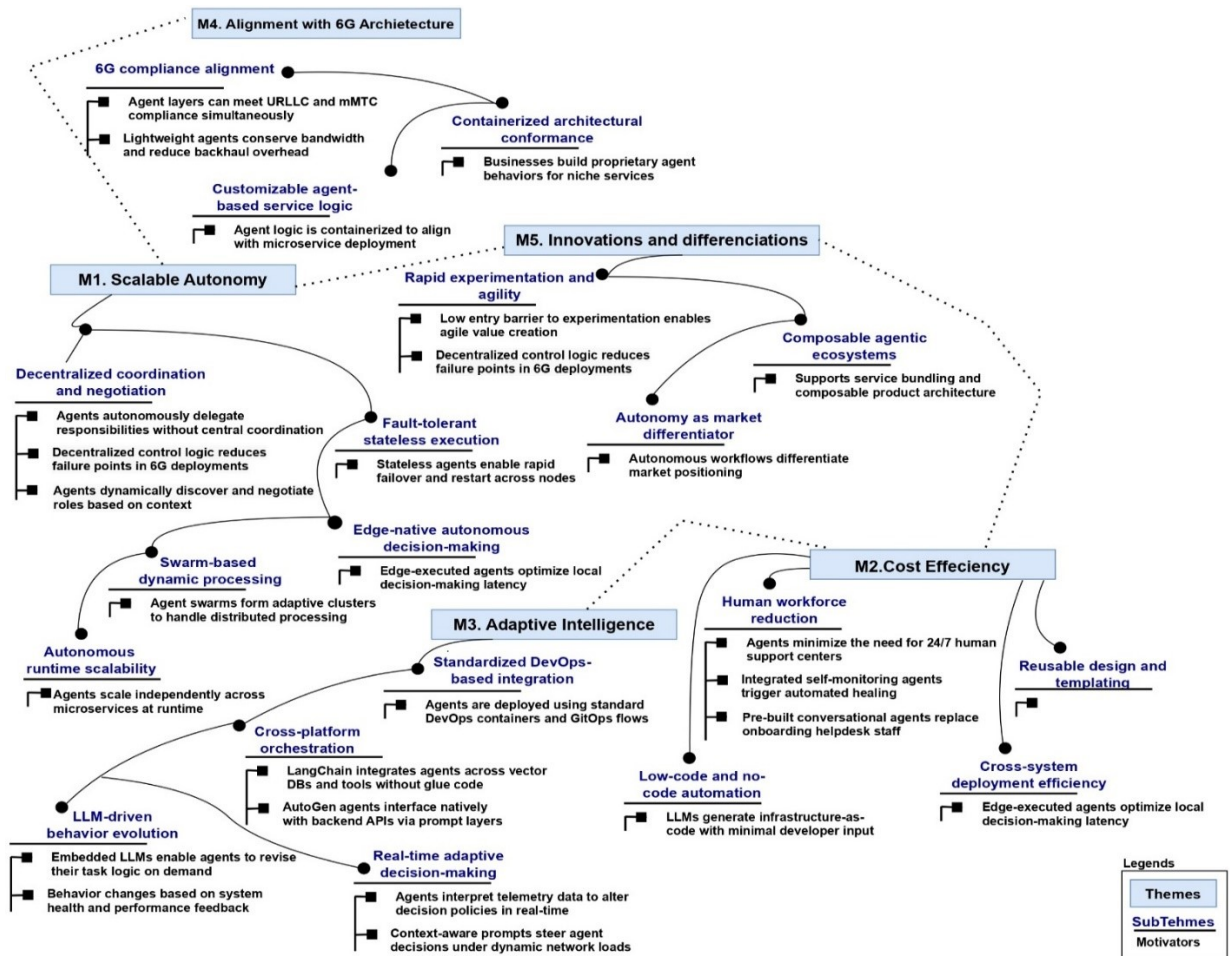

Fig. 2. Agentic Ai In 6g Software:Motivators .

These concerns are compounded by security and quality assurance gaps, as open-source agent-runtimes often lack thorough audit mechanisms, and LLMs introduce unpredictable, non-deterministic behaviors[24, 30]. Furthermore, underspecified state handling such as vague memory management practices[8, 19] and the scarcity of large-scale validation test suites contributes to fragile and untrustworthy system performance[8, 20, 30].

**D2: Trust and Accountability**. Due to the probabilistic nature of LLM-based decision-making, many deployments suffer from explainability and auditability gaps[25, 39], making it difficult to trace logic flows[19, 36], detect faults[20], or provide justifications to stakeholders[24, 36]. Compounding this issue is a governance void in autonomous systems[30], where responsibility is fragmented and rollback paths for faulty agent actions are often absent[10, 30]. Such opacity in behavioral disclosure[10] especially when agents operate across domains raises red flags in sectors requiring compliance and ethical oversight[5, 36].

**D3: Integration Complexity** captures the architectural strain of embedding agents within legacy software stacks and distributed environments[4, 7, 8]. Many organizations face legacy system incompatibility[25], where agents demand major refactoring and may not align with real-time system requirements[5, 31]. Beyond that, the need for high coupling with backend logic[20, 22] such as state handlers or tightly bound APIs[27] undermines modularity and limits reuse[4]. Simultaneously, debugging and observability constraints arise due to the distributed and asynchronous nature of multi-agent interactions, for which current tooling is often inadequate[17, 18].

**D4: Organizational Readiness** reflecting the socio-technical preparedness of teams to operationalize agentic systems. Many organizations exhibit a lack of internal skills and structures[27], with limited expertise in areas such as prompt chaining or orchestration design[5, 10], and blurred team responsibilities for supervising autonomous workflows[19]. Additionally, cultural and political resistance persists within traditional DevOps or system administration groups[5, 7, 31], often driven by fear of displacement or paradigm fatigue[38]. These hurdles are exacerbated by value measurement and uncertainty[7], as businesses struggle to quantify agent-driven ROI and rarely possess governance frameworks tailored for agent oversight[36, 38, 40].

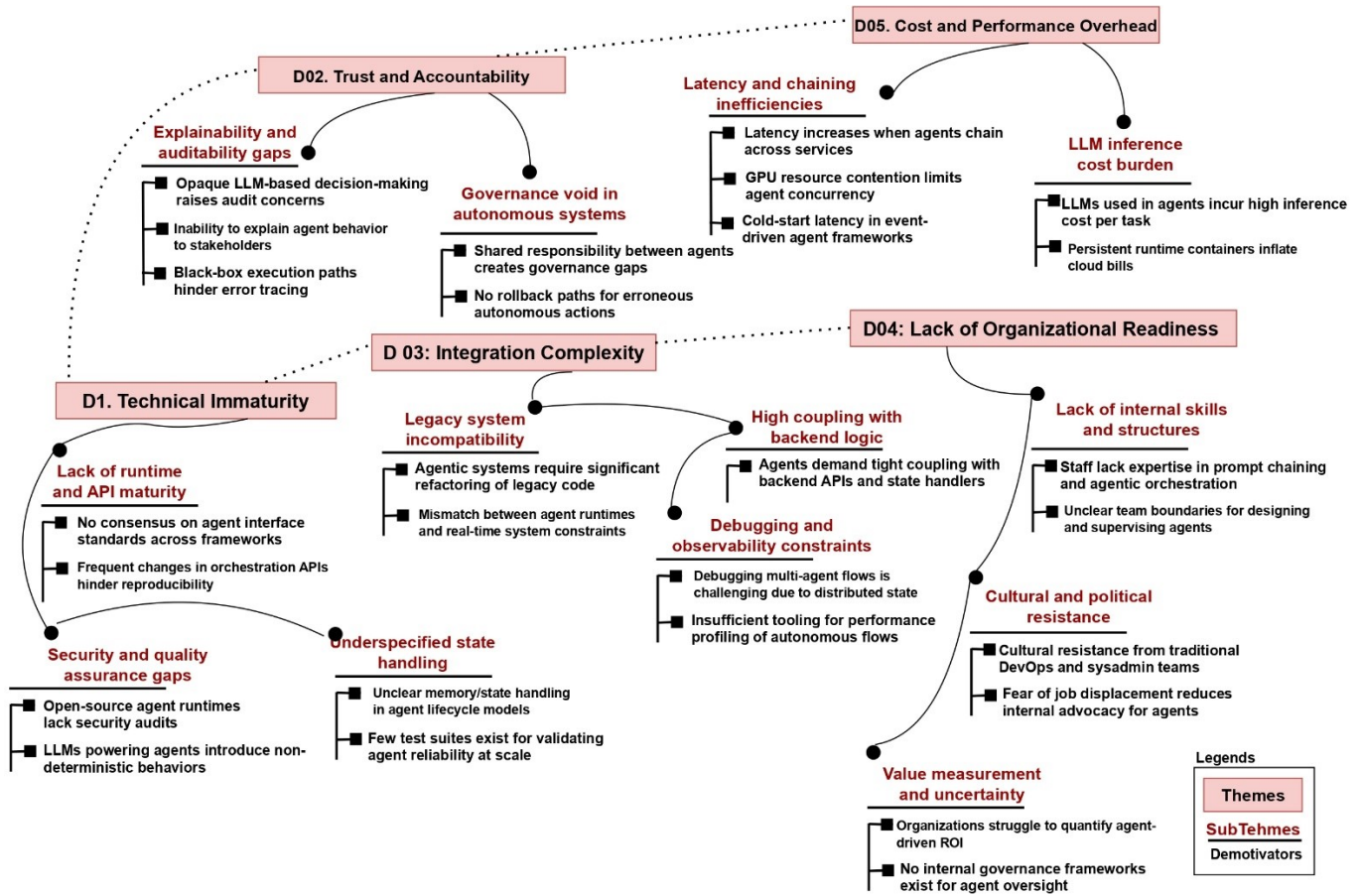

Fig. 3. Agentic Ai In 6g Software:Demotivators

Finally, **D5: Cost and Performance** Overheads serve as tangible deterrents. Agentic deployments powered by LLMs often suffer from a LLM inference cost burden[24, 38], as high per-task inference loads, and long-lived containers inflate operational costs[14, 31, 38]. In performance-sensitive settings, latency[2, 22] and chaining inefficiencies emerge when agents chain across multiple services or contend for GPU resources, leading to cold-start delays and degraded responsiveness[18, 19]. These performance constraints raise broader concerns around scalability bottlenecks, particularly in use cases where ultra-reliability and low latency are paramount.

In summary, while agentic software offers significant opportunities for advancing 6G ecosystems, its adoption is shaped by both strong motivators and critical demotivators. Realizing its full potential will require addressing technical, organizational, and economic concerns, alongside fostering architectural maturity, operational readiness, cultural alignment, and effective governance.

## IV. PROPOSED MATURITY MODEL AND RESEARCH CONTRIBUTION

The primary objective of this research is to develop an Agentic AI Software Engineering Maturity Model (AAISEMM) tailored for 6G software businesses. This model is intended to support software developers and organizations in evaluating, enhancing, and structuring their agentic AI software processes effectively and efficiently within 6G environments. AAISEMM will serve as a strategic tool to help software experts identify and analyze key challenges, adopt relevant best practices, and guide the evolution of agent-driven architectures across the Data, Business Logic, and Presentation layers.

This study is original in its focus on bridging agentic AI and software engineering maturity within the 6G context—a critical yet underexplored area. To our knowledge, no prior work has systematically addressed the readiness, process improvement, or strategic adoption of agentic AI software systems in 6G-oriented business environments. This proposed model contributes both theoretically and practically by offering a structured approach for organizations to navigate the transition toward agentic-first software development.

## ACKNOWLEDGMENT

This research was funded by the Business Finland project 6G Bridge:6G software for extremely distributed and heterogeneous massive networks of connected devices (8516/31/2022). The authors also acknowledge the use of generative AI tools, including ChatGPT, and the writing assistant tool Grammarly, which were instrumental in addressing writing issues in this paper. The authors reviewed and substantially modified all AI-generated content and assumed full responsibility for the final content.

**Appendix:** The list of selected studies used in this research is available at the following link:
https://doi.org/10.5281/zenodo.16743871